\documentstyle[12pt,worldsci]{article}

\def\nn{\nonumber}

\def\MeV{{\rm \ MeV}}
\def\sss{\scriptscriptstyle}

\def\ngb{Nambu-Goldstone boson}

\def\G{\Gamma}
\def\pf{p_{\sss F}}

\def\wf{w_{\sss F}}
\def\wgt{w_{\sss GT}}
\def\GF{G_{\sss F}}

\def\bb{\beta\beta}
\def\bbtn{\bb_{2\nu}}
\def\bbm{\bb_{\varphi}}

\def\bbzn{\bb_{0\nu}}

\def\etal{{\it et al.}}

\def\gtwid{\mathrel{\raise.3ex\hbox{$>$\kern-.75em\lower1ex\hbox{$\sim$}}}}
\def\ltwid{\mathrel{\raise.3ex\hbox{$<$\kern-.75em\lower1ex\hbox{$\sim$}}}}

\def\frac#1#2{{\scriptstyle{#1 \over #2}}}
\def\slp{{\raise.15ex\hbox{$/$}\kern-.57em\hbox{$\partial$}}}
\def\darr{\raise1.5ex\hbox{$\leftrightarrow$}\mkern-16.5mu \slp}
\def\gv{g_{\sss V}}
\def\ga{g_{\sss A}}

\def\bfp{{\bf p}}

\def\bfrnm{{\bf r}_{\!nm}}

\def\ellsl{{\raise.15ex\hbox{$/$}\kern-.57em\hbox{$\ell$}}}
\def\psl{{\raise.15ex\hbox{$/$}\kern-.57em\hbox{$p$}}}
\def\qsl{{\raise.15ex\hbox{$/$}\kern-.57em\hbox{$q$}}}

\def\Nd{$^{150}$Nd}
\def\Mo{$^{100}$Mo}
\def\Se{$^{82}$Se}
\def\Nd{$^{150}$Nd}

\def\ol#1{\overline{#1}}

\def\roughlyup#1{\mathrel{\raise.3ex\hbox{$\sim$\kern-.75em
\lower1ex\hbox{$#1$}}}}
\def\roughlydown#1{\mathrel{\raise.3ex\hbox{$#1$\kern-.75em
\lower1ex\hbox{$\sim$}}}}
\def\bra{\langle}
\def\ket{\rangle}

\def\simeq{\roughlyup-}

\def\dket{\rangle\!\rangle}
\def\dbra{\langle\!\langle}

\def\eqa{\begin{eqnarray}}
\def\eeqa{\end{eqnarray}}
\def\eq{\begin{equation}}
\def\eeq{\end{equation}}
\def\Sca{{\cal A}}

\def\Sco{{\cal O}}

\def\bfp{{\bf p}}

\begin{document}

\rightline{January 1994}
\rightline{McGill-94/06, NEIP-94-002}
\rightline{TPI-MINN-94/3-T, UMN-TH-1236/94}
\rightline{hep-ph/9401334}

\title{{\bf WHAT NEW PHYSICS CAN DOUBLE BETA DECAY \\
EXPERIMENTS HOPE TO SEE? }}
\author{C.P. BURGESS\thanks{Talk presented to the International 
Conference on Non-Accelerator Particle Physics, Bangalore India, 
January 1994.}\\
{\em Institut de Physique, Universit\'e de Neuch\^atel \\
1 Rue A.L. Breguet, CH-2000 Neuch\^atel, Switzerland.}\\
\vspace{0.3cm}
and\\
\vspace*{0.3cm}
JAMES M. CLINE\\
{\em Theoretical Physics Institute, The University of Minnesota \\
Minneapolis, MN, 55455, USA.}\\
}
\maketitle
\setlength{\baselineskip}{2.6ex}

\begin{center}
\parbox{13.0cm}
{\begin{center} ABSTRACT \end{center}
{\small \hspace*{0.3cm}
Double-beta decay experiments have been traditionally interpreted in 
terms of the Gelmini-Roncadelli triplet majoron model which has 
since been ruled out by the LEP data on the $Z$ resonance. We 
therefore systematically re-examine the kinds of physics to which 
double-beta decay experiments might be sensitive, with particular 
attention paid to a potential scalar-emitting mode. We find six
broad categories of models, including some new categories which have 
not been previously considered. Models in these new classes robustly differ 
from the old ones in the electron energy spectrum that they predict, and 
depend on different nuclear matrix elements.  For models in which 
the electron neutrino mixes with sterile neutrinos, an observable 
double-beta decay signal typically implies a sterile-neutrino mass
in the neighbourhood of 1 MeV to 1 GeV.
}}
\end{center}

\section{Introduction}

Over the past five years double-beta-decay experiments have come of age. Since
the first direct observation of Standard-Model neutrino-emitting double beta
decay ($\bbtn$) was made several years ago \cite{first}, other experimenters
have taken up the challenge and have observed this decay in several elements.

Much of the original motivation for these experiments was not so much to find
these expected Standard-Model decays, but rather to search for nonstandard
$\bbzn$ decays in which the two outgoing electrons are unaccompanied 
by neutrinos. Such a decay must violate the conservation of electron 
number, $L_e$, and as such would be a smoking gun for `new physics' 
from beyond the Standard Model. The possibility
of breaking electron number spontaneously also motivated searching for a third
decay, $\bbm$, in which the electrons emerge together with the appropriate
\ngb, called the majoron.  Both of these decays were indeed predicted
\cite{georgi}${}^,$ \cite{Doi} by a simple and elegant  model, due to Gelmini
and Roncadelli \cite{GR}, in which lepton number was spontaneously broken by an
electroweak-triplet Higgs field. Although this model has since been ruled out
by the observations at the $Z$ resonance at LEP,  it has still remained as 
the paradigm against which double-beta-decay experiments compare 
their results.

Interest in understanding the kind of physics to which these experiments can be
sensitive was recently revived by the tentative observation of an excess of
electrons near, but below, the endpoint for the decays of \Mo, \Se\ and \Nd, 
with a statistical significance of 5$\sigma$ \cite{Moe}. This observation 
echoed earlier indications for such an excess in the decay of $^{76}$Ge 
\cite{Avignone}, although this earlier evidence was
later ruled out both by the initial investigators, as well as by others
\cite{nomajoron}.
Hints of excess events also persisted in the ${}^{76}$Ge data
\cite{ge}${}^,$ \cite{priv}${}$\hspace{-1mm}, although at a tenth or less of
the originally-detected rate.  Although, at present,
most of the anomalous events reported by the Irvine group seem to be due to
resolution problems for  the higher-energy electrons \cite{MoeII},  
there remains a smaller set of residual events whose magnitude is 
consistent with observations from other
experiments.\cite{heidelberg}${}^,$\cite{neuchatel} ${}$

This experimental activity has provoked a theoretical re-examination
\cite{berezhiani}${}^,$\cite{ourletter}${
}^,$\cite{carone}${}^,$\cite{ourpaper}\ of the
kinds of new physics that could be expected to be detectable with the current
sensitivity of $\bb$ experiments. In particular, attention has been devoted to
understanding the implications for these experiments that can be extracted from
the spectacular experiments at LEP.  The principal idea is to evade the LEP
bounds by forbidding any coupling between the $Z$ boson and any new 
light degrees of freedom which appear in the model. This
is most naturally ensured by making all such new particles electroweak
singlets, generalizing the old singlet-majoron model\cite{CMP}.  
The purpose of this article is to summarize  the results of this re-examination.
The general conclusion can be encapsulated by
the statement that double-beta decay can be generated at an observable level,
but only if the new physics has rather different properties than have
previously been assumed.

\section{General Properties of Double Beta Decay}

For the purposes of classification it is convenient to write the rates for
double-beta decay in the following 
way:\cite{ourletter}${}^,$\cite{ourpaper}
\eq
\label{genericrate}
d\G(\bb_i) = {(\GF\cos\theta_{\sss C})^4 \over 4\pi^3} \; \left| \Sca(\bb_i)
\right|^2 d\Omega(\bb_i),
\eeq
where $\GF$ is the Fermi constant, $\theta_{\sss C}$ the Cabibbo angle,
$\Sca(\bb_i)$ a nuclear matrix element, and $d\Omega(\bb_i)$ the differential
phase space for the particular process. The index `$i$', in
$\bb_i$ represents the possible decays $\bbtn$, $\bbzn$, $\bbm$ {\it etc}.

{}From eq.~(\ref{genericrate}) one can see there are two quantities to which
double-beta-decay experiments are sensitive. They are:

\bigskip\noindent
{\bf 1. The Electron Energy Spectrum:}
This quantity is the relative frequency of the observed outgoing electrons, as
a function of their energies, $\epsilon_k$ ($k=1,2)$. To a good 
approximation (a few percent) the {\it shape} of this distribution is 
completely described by the factor, $d\Omega(\bb_i)$, of 
eq.~(\ref{genericrate}), and is therefore completely independent
of the uncertainties that are associated with the nuclear matrix elements.
This is because the maximum energy, $Q\sim (1-3) \MeV$, that is released by the
decay is much smaller than the typical momentum transfer, $\pf \sim 100 \MeV$,
between the decaying nucleons that sets the scale for the momentum dependence
of the nuclear matrix elements.

For  $\bbzn$ decay, $d\Omega(\bbzn)$ is given by
\eq
\label{bbznphsp}
d\Omega(\bbzn) = {1\over 64\pi^2} \; \delta(Q - \epsilon_1-\epsilon_2)
\prod_{k=1}^2 p_k \epsilon_k F(\epsilon_k) \; d\epsilon_k.
\eeq
Here $p_k = |\bfp_k|$ is the magnitude of the electron
three-momentum, and the endpoint energy, $Q$, for the electron spectrum
is given in terms of the energies of the initial and final nuclear states, $M$
and $M'$, and the electron mass, $m_e$, by $Q = M-M' - 2m_e$.  Finally,
$F(\epsilon)$ is the Fermi function, normalized to unity in the limit of 
vanishing nuclear charge. The corresponding quantity for the other 
processes has  a similar form,
\eq
\label{phsp}
d\Omega(\bb_i) = {1\over 64\pi^2} \; (Q - \epsilon_1-\epsilon_2)^{n_i}
   \prod_{k=1}^2 p_k \epsilon_k F(\epsilon_k) \;  d\epsilon_k.
\eeq
(The above formula applies to the scalar-emitting decays provided that the
emitted boson is massless. Should it have mass $m$ then the factor 
$(Q -\epsilon_1 - \epsilon_2)$
should be replaced by  $((Q -\epsilon_1 - \epsilon_2)^2 -  m^2)^{1/2}$.)

It is only the {\it spectral index}, $n_i$, which differs depending on the type
of decay, and whose implications for the spectral shape are detectable
experimentally. The standard $\bbtn$ decay has $n_{2\nu}=5$, and 
the resulting spectral shape may be compared with the cases $n=1$ and 
$n=3$, which arise in all other practical examples, in Fig. 1.

\vspace{0.08in}
\begin{center}
\let\picnaturalsize=N
\def\picsize{2.3in}
\def\picfilename{spectrum.ps}
\ifx\nopictures Y\else{\ifx\epsfloaded Y\else\input epsf \fi
\let\epsfloaded=Y
\centerline{\ifx\picnaturalsize N\epsfxsize \picsize\fi \epsfbox{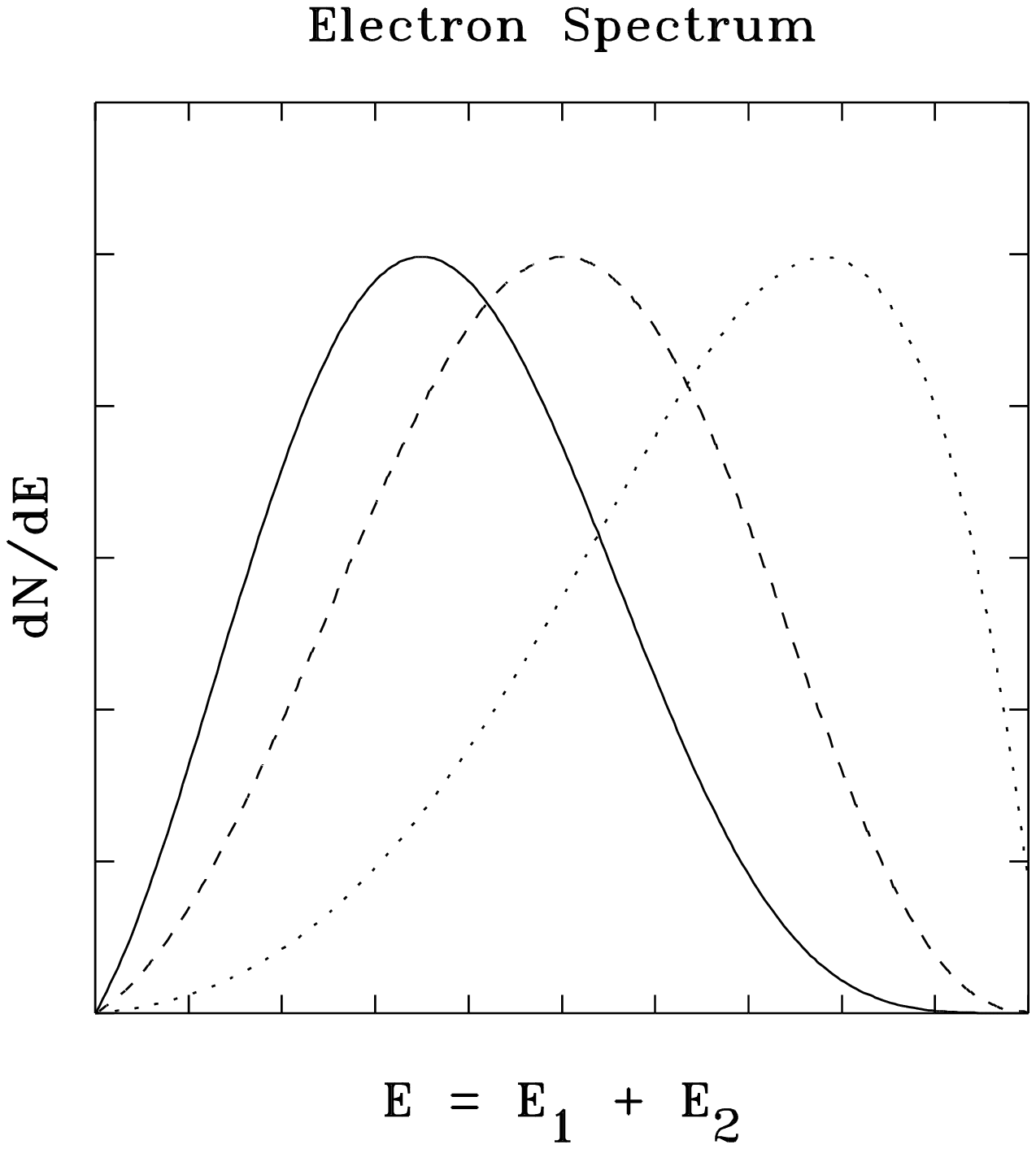}}}\fi

\medskip
{\bf Figure 1}\\
\medskip
{\sl The electron energy spectrum corresponding to the spectral indices\\
$n=1$ (dotted), $n=3$ (dashed) and $n=5$ (solid).}
\end{center}

\bigskip\noindent
{\bf 2. The Integrated Rate:}
The other observable quantity is the {\it normalization}, $\Sca(\bb_i)$,  of
the spectrum, as determined by the total rate for each of the possible 
types of decays. It is here that one encounters the uncertainties 
associated with calculating nuclear matrix elements.  It is convenient to parameterize our ignorance of these matrix elements by writing them 
in terms of a model-independent set of form factors.
The basic nuclear matrix element which determines the double-beta decay rates
is\cite{ourpaper}
\eq
\label{matrixelement}
W_{\alpha\beta}(p) \equiv (2 \pi)^3 \, \sqrt{ { E E'\over M M'}} \;
\int d^4x \;\bra N'|T^* \left[ J_\alpha(x) J_\beta(0) \right] | N\ket \;
e^{ipx},\eeq
where $J_\mu = \bar{u} \gamma_\mu (1+\gamma_5) d$ is the weak charged
current that causes transitions from neutrons to protons, and $|N \ket$ and
$|N' \ket$ represent the initial and final $0^+$ nuclei in the decay. $E$ and
$M$ are the energy and mass of the initial nucleus, $N$, while $E'$ and $M'$
are the corresponding properties for the final nucleus, $N'$. For the decays 
of interest the most general possible form for $W_{\alpha\beta}$ 
is \cite{ourpaper}:
\eqa
\label{formfactors}
W_{\alpha\beta}(p) &=& w_1 \; \eta_{\alpha\beta} + w_2 \; u_\alpha
        u_\beta +w_3 \; p_\alpha p_\beta + w_4 \; (p_\alpha u_\beta +
                                        p_\beta u_\alpha) \nn\\
    &&+ w_5 \; (p_\alpha u_\beta - p_\beta u_\alpha) +
     iw_6 \; \epsilon_{\alpha\beta\sigma\rho}
     u^\sigma p^\rho ,
\eeqa
where $u_\alpha$ is the four-velocity of the initial and final nucleus, and
the six Lorentz-invariant form factors, $w_a = w_a(u\cdot p, p^2)$, are
functions of the two independent invariants that can be constructed from
$p_\mu$ and $u_\mu$.

Since the literature --- for which there are a number of excellent reviews
\cite{DoiTomoda}${}^,$\cite{reviews} --- tends to quote expressions in which
the nuclear matrix elements have been evaluated in a particular model of the
nucleus, it is useful to have expressions for these form factors using these 
models. For instance $\bbtn$, $\bbzn$ and some kinds of $\bbm$ 
decays involve only the combination ${W^\alpha}_\alpha$, which 
may be written using the closure and nonrelativistic impulse 
approximations\cite{ourpaper} as ${W^\alpha}_\alpha = \wf - \wgt$,
with:
\eqa
\label{connection}
\wf &=& {2i \mu \gv^2\over p_0^2 - \mu^2 + i\varepsilon}
\; \dbra N'|\sum_{nm}e^{-i\bfp\cdot\bfrnm}\tau^+_n\tau^+_m |N \dket;\nn\\
         \wgt &= &{2i \mu \ga^2\over p_0^2 - \mu^2 + i\varepsilon} \;
\dbra N'|\sum_{nm} e^{-i\bfp \cdot \bfrnm} \tau^+_n\tau^+_m
        \; \vec\sigma_n \! \cdot\! \vec\sigma_m |N \dket .
 \eeqa
Here $\mu \equiv \ol{E} - M$ is the average excitation energy of the
intermediate nuclear state, $\bfrnm$ is the separation in position between
the two decaying nucleons, $\gv \simeq 1$ and $\ga \simeq 1.25$ are the
vector and axial couplings of the nucleon to the weak currents, and
$\dbra N'| \Sco |N \dket$ represents a reduced matrix element from which
the nuclear centre-of-mass motion has been extracted.

\section{A General Classification}

Given the limited number of observables which are measured in double beta
decay models, there are essentially two questions which qualitatively
distinguish the signatures of all models which can produce observable 
$\bb$ decay.  These are:
\begin{description}
\item[Q1:]
Is electron number, $L_e$, broken?
\item[Q2:]
Are there light bosons in the model which can produce $\bbm$ decays?
\end{description}
If the answer to this second question should be `yes', then two more questions
are needed to distinguish the possibilities for $\bbm$ decay:
\begin{description}
\item[Q2a:]
Is the light particle a Goldstone boson?
\item[Q2b:]
What are the light boson's quantum numbers if electron number is conserved?
\end{description}
We consider here only the case of a light scalar boson, although a similar
analysis for a vector particle follows similar lines\cite{carone}.

The implications of these questions to the two types of experimental
signatures that are possible is summarized in Table I.

\bigskip
\begin{center}
\begin{tabular}{cccccc}
\noalign{\hrule}\noalign{\smallskip}\noalign{\hrule}\noalign{\medskip}
& $L_e$ & A New Scalar: & $\bbzn$ & $\bbm$ & Spectral Index \\
\noalign{\medskip}\noalign{\hrule}\noalign{\smallskip}
\noalign{\hrule}\noalign{\medskip}
IA & Broken & Does Not Exist & Yes & No & N.A. \\
IB & Broken & Is Not a Goldstone Boson & Yes & Yes & $n=1$ \\
IC & Broken & Is a Goldstone Boson & Yes & Yes & $n=1$ \\
\noalign{\medskip}\noalign{\hrule}\noalign{\medskip}
IIA & Unbroken & Does Not Exist & No & No & N.A. \\
IIB & Unbroken & Is Not a Goldstone Boson ($L_e=-2$) & No & Yes & $n = 1$ \\
IIC & Unbroken & Is Not a Goldstone Boson ($L_e=-1$) & No & Yes & $n = 3$ \\
IID & Unbroken & Is a Goldstone boson ($L_e=-2$) & No & Yes & $n=3$ \\
IIE & Unbroken & Is a Goldstone boson ($L_e=-1$) & No & Yes & $n=5$ \\
\noalign{\medskip}\noalign{\hrule}\noalign{\medskip}
\end{tabular}

{\bf Table I }\\
{\sl A list of alternatives for modelling double beta decay.}
\end{center}
\bigskip

Six broad categories of models emerge from an inspection of Table I.

\begin{enumerate}
\item
The most conservative option is the first category --- case IIA of the 
Table --- which predicts no new physics to be seen in $\bb$ 
experiments.
\item
The next most conservative case is case IA, which is distinguished by a
potential $\bbzn$ signal but absolutely no scalar-emitting decays. This 
implies the standard $\bbtn$ electron spectrum away from the endpoint.
\item
The next category contains two classes of models which can be indistinguishable
from the point of view of $\bb$ experiments --- IB and IC of the Table. 
(The only way these could be distinguished would be if, in case IB, the scalar
mass were nonzero and appreciable in comparison to the electron endpoint
energy, $Q$. ) 
This class --- which includes the old Gelmini-Roncadelli model ---
predicts\cite{berezhiani}${}^,$\cite{ourletter}${}^,$\cite{ourpaper}\
the standard GR form for the electron spectrum in scalar decay ({\it i.e.} 
it has spectral index $n=1$, and depends only on the matrix elements 
${W^\alpha}_\alpha$\cite{ourpaper}).
\item
Case IIB of the table forms another class all by itself. It would only be
clearly distinguished from cases IB and IC if $\bbzn$ decay should be 
found to be nonzero, since this is absolutely forbidden in case IIB. 
If the scalar-emitting decay, $\bbm$, should be detected without 
finding an accompanying $\bbzn$ signal, then (for effectively 
massless scalars) cases IB, IC and IIB could not be distinguished 
simply by looking at the electron spectrum.

This is, at first sight, surprising since these categories differ in whether
they break electron number, and in whether the light scalar is a Goldstone 
boson or not. After all, Goldstone bosons couple derivatively and this 
might be expected to be reflected in the predicted electron spectrum. 
The main point here\cite{ourletter}${}^,$\cite{ourpaper} is that the 
present detection limit for $\bbzn$ is so strong that it
forces all of the $L_e$-breaking terms in cases IB and IC to be so small
as to be irrelevant for $\bb$ decay. As a result, so far as the $\bbm$ signal
is concerned, the predictions of the $L_e$-breaking models of cases IB and 
IC are for all practical purposes indistinguishable from those of the
$L_e$-preserving models in IIB.  They also depend on the usual matrix 
elements ${W^\alpha}_\alpha$.\cite{ourpaper}
\item
Next comes cases IIC and IID --- models which are identical in
their implications for $\bb$ experiments. They both predict an electron
spectrum which is qualitatively different from that of the older GR 
model, producing electrons which are softer than those of the GR 
majoron-emitting decay, but which are harder
than those of the Standard-Model $\bbtn$ decay.

The reasons for this alternative spectrum differs for cases IIC and IID. For
case IID, conservation of $L_e$ and the condition that the emitted scalar 
be a Goldstone boson imply\cite{ourletter}${}^,$\cite{ourpaper}\ that 
the scalar emission amplitude must be proportional to the scalar energy, 
and this implies the observed softening of the electron spectrum. (Models 
in this class also depend on different nuclear matrix elements, depending 
as they do on the form factors $w_5$ and $w_6$.\cite{ourpaper}) For 
case IIC conservation of electron number requires the emission of 
{\it two}  scalars at a time. The additional scalar phase space is then 
responsible for the additional suppression of high-energy
electrons. Although the original models of this type\cite{twomajoron}\ are
ruled out by the LEP data, alternative singlet-type models are also possible
\cite{twosinglets}.
\item
The final category is class IIE, for which $L_e$ is
conserved, the light scalars are Goldstone bosons, and for which $L_e = -1$.
In this case the spectrum is expected to be softened compared to the GR
model by ($i$) two powers of  $(Q-E)$, since the emitted particles are 
Goldstone bosons, and ($ii$) by two additional powers due to
phase space since $L_e$ conservation requires two bosons to be emitted at a 
time. The spectral index for this spectrum is therefore expected to be $n=5$,
although no models of this type have yet been constructed. This would make
it indistinguishable in shape from the standard $\bbtn$ decay.
\end{enumerate}

\section{More Detailed Predictions}

In order to know whether all of the options given in Table I are actually
viable, it is necessary to compute representative models in each category.
Only then is it possible to check that the properties that are required for an
observable $\bb$ signal are consistent with all other neutrino bounds.
Prominent among these bounds is consistency with Big-Bang 
Nucleosynthesis, since any model with a light scalar coupling significantly 
to neutrinos can easily ruin the present understanding of the primordial 
origin of the light elements. Although such an analysis has been 
done\cite{berezhiani}${}^,$\cite{ourpaper}
for models in categories IB, IC, IIB and IID, work is still in progress 
for cases IIC and IIE.

Some general features do emerge from these analyses, however. For instance,
a very broad category of models introduces the coupling between the light
scalar and the electron neutrino by mixing $\nu_e$ with various species of
sterile neutrinos, $\nu_s$. A general feature of all such models is the
necessity of having a neutrino state with a mass at least as large as 
$\sim 1$ MeV if an observable exotic $\bbm$ rate is 
required.\cite{ourpaper} This is because in such models the
$\bbm$ rate vanishes in the limit that all neutrinos are degenerate, since in
this limit there need be no mixing between electroweak eigenstates. This
implies, in particular, that for light neutrinos, the $\bbm$ decay rate
is always suppressed by explicit factors of neutrino masses divided by the
nuclear-physics scales, $\pf \sim 100$ MeV, that are relevant to $\bb$ decay.
An experimentally observable rate therefore requires at least some 
neutrino masses that are at least of order a few MeV.

Another general feature concerns the understanding of why the light degrees of
freedom in the model should be so light. In this regard models in classes IB,
IIB and IIC are theoretically unattractive in that they must build in a light
scalar mass completely by hand. (An equally small scale is also 
required for models in class IC, since although here the Goldstone 
boson is naturally massless, the absence
of $\bbzn$ decay requires a similarly small fine-tuning of the scale of $L_e$
breaking.) In all of these models such a small scale is typically only possible
if the scalar potential involves dimensionless self-couplings that are as small
as $10^{-14}$. It is nevertheless sometimes possible to make these 
models natural in the technical sense.\cite{ourpaper}${}^,$\cite{common} 
Models in category IID (and presumably IIE) are much more appealling 
in this way since for them the scale of symmetry breaking can be 
orders of magnitude higher.

\section{Conclusions}

It is clear that $\bb$ experiments do provide a window on potential new
physics, and in a way which fundamentally probes the 
validity of electron-number conservation. Viable models 
exist which can account for detectable signals in
these experiments, without being in conflict with other data, such as the
properties of the $Z$ boson as measured at LEP. Most interestingly, however,
the properties of these models can differ significantly from what would be
expected based on the early model building of previous decades. In particular,
perhaps the theoretically best-motivated new models predict an entirely
different electron energy spectrum, which should be searched for in the
data of the ongoing experiments.

\section{Acknowledgements}

C.B. would like to thank the organizers for their warm
hospitality throughout the conference.
This research received support from the Swiss National
Foundation, N.S.E.R.C.\ of Canada, les Fonds F.C.A.R.\ du Qu\'ebec, 
and DOE grant DE-AC02-83ER-40105.

\def\pr#1{\it Phys.~Rev.~{\bf #1}}
\def\np#1{\it Nucl.~Phys.~{\bf #1}}
\def\pl#1{\it Phys.~Lett.~{\bf #1}}
\def\prc#1#2#3{{\it Phys.~Rev.~}{\bf C#1} (19#2) #3}
\def\prd#1#2#3{{\it Phys.~Rev.~}{\bf D#1} (19#2) #3}
\def\prl#1#2#3{{\it Phys. Rev. Lett.} {\bf #1} (19#2) #3}
\def\plb#1#2#3{{\it Phys. Lett.} {\bf B#1} (19#2) #3}
\def\npb#1#2#3{{\it Nuc. Phys.} {\bf B#1} (19#2) #3}
\def\etal{{\it et.al. \/}}

\bibliographystyle{unsrt}

\end{document}